\documentclass[a4paper]{jpconf}
\usepackage{graphicx}
\begin{document}
\title{Where are the trapped surfaces?}

\author{Jan E \AA man$^1$, Ingemar Bengtsson$^1$, and Jos\'e M M Senovilla$^2$}

\address{$^1$ Fysikum, Stockholms Universitet, S-106 91 Stockholm, Sweden}

\address{$^2$ F\'isica Te\'orica, Universidad del Pa\'is Vasco, Apartado 644,
48080 Bilbao, Spain}

\ead{ingemar@physto.se}

\begin{abstract}
We discuss the boundary of the spacetime region through each point 
of which a trapped surface passes, first in some simple soluble examples, 
and then in the self-similar Vaidya solution. For the latter 
the boundary must lie strictly inside the event horizon. We present 
a class of closed trapped surfaces extending strictly outside the apparent 
horizon. 
\end{abstract}

\section{Introduction}
Roger Penrose tells the story of how the importance of closed trapped 
surfaces occurred to him while crossing a street in London. He then promptly 
forgot about the matter, but---fortunately---recalled it later in the day 
\cite{Penrose}. One can make a case for trapped surfaces being the 
single most important ingredient in gravitational collapse. Their popular acclaim 
has been clouded by the fact that they remain hidden behind the 
event horizon, but numerical relativists do encounter them, while 
the very existence of the event horizon may be in some doubt from quantum 
gravity. 

Can one {\it define} a black hole using its trapped surfaces \cite{Hayward}? An 
important notion here is the dynamical horizon, a 
spacelike hypersurface foliated by marginally trapped surfaces \cite{Ashtekar}. 
But not enough is known about the uniqueness of such objects \cite{Greg}. 
Indeed little is 
actually known about where trapped surfaces occur even in the simple 
Vaidya solution, which  describes the collapse of a spherically 
symmetric cloud of incoherent radiation (or ``null dust'') \cite{Vaidya}. 
This is the situation that we have tried to remedy, with partial success.  

Any dynamical horizon lies inside the trapping boundary, defined as the 
boundary of an inextendible spacetime region 
for which each point lies on some trapped surface \cite{Hayward}. In 
very simple cases this boundary is easily 
found. In the Vaidya spacetime it must lie somewhere between the event horizon 
and a spherically symmetric dynamical horizon, but coincides with neither 
\cite{IBJS, JS}. Here we will present a family of closed trapped surfaces 
that extend into the no-man's land between them.
The surfaces we consider are genuinely trapped, in that both of their null 
expansions are non-positive, whereas much of the recent literature 
concerns outer trapped surfaces for which no condition is imposed 
on the inner expansion. There are good reasons 
for this \cite{Lars}, but there are reasons for our choice too 
\cite{Hayward, Ashtekar}. The reader must not confuse the two. 
Eardley conjectured that the boundary of the region through which 
closed outer trapped surfaces pass is the event horizon 
\cite{Eardley}, and for the Vaidya case this was proven by Ben-Dov 
\cite{BenDov}. 

\section{Simple examples and how to address them}
A spacetime where the trapping 
boundary can be identified almost by inspection is the recollapsing 
$k = 1$ Friedmann model. It is foliated by round 3-spheres and has a moment 
of time symmetry for which the trace $K$ of the second fundamental form 
vanishes. Any equatorial sphere in such a 3-sphere is minimal within the 
3-sphere, and trapped if $K < 0$. The null expansions vanish 
at the moment of time symmetry, so we expect this to be the 
trapping boundary. But how do we prove that a closed trapped 
surface cannot extend partly below it?  

Let $e^a_i$ project the cotangent space onto the cotangent space of some 
spacelike surface of codimension two. Let $\bar{\xi}^i$ be the projection of 
an arbitrary spacetime vector field $\xi^a$, and let $K_{ij}^{\ \ a}$ be 
the shape tensor of the surface. Then 

\begin{equation} e^a_ie^b_j\nabla_a\xi_b = \bar{\nabla}_i\bar{\xi}_j + 
K_{ij}^{\ \ a}\xi_a \ . \end{equation} 

\noindent The derivative on the right hand side uses the Levi-Civita 
connection of the first fundamental form. If we symmetrise and then 
contract with the first fundamental form we obtain

\begin{equation} \frac{1}{2}\gamma^{ij}e^a_ie^b_j{\cal L}_{\xi}g_{ab} = 
\bar{\nabla}_i\bar{\xi}^i + H^a\xi_a \ , \label{ojoj} \end{equation}

\noindent where $H^a = \gamma^{ij}K_{ij}^{\ \ a}$ is the mean curvature 
vector. This is future directed and timelike for a trapped surface. 
Suppose that the vector field $\xi^a$ is future directed and timelike too, 
so that $H^a\xi_a < 0$, and suppose that $\xi^a$ is such that the left 
hand side of this equation is non-negative. If we integrate over a closed 
surface the divergence goes away and we have a contradiction. Either the 
surface extends into a region where $\xi^a$ is spacelike, or else it is not 
trapped \cite{Seno}. 

If the vector field $\xi^a$ is hypersurface forming we can say more. Assume 
that there exist functions $F > 0$ and $\tau$ such that 

\begin{equation} \xi_a = - F\nabla_a \tau \ . \label{tau} \end{equation}

\noindent On a closed surface the function $\tau$ will assume a minimum. 
At that minimum 

\begin{equation} \bar{\nabla}_i\xi^i = - \bar{\nabla}_i(F \bar{\nabla}^i\tau) = 
- F \bar{\nabla}_i\bar{\nabla}^i\tau - \bar{\nabla}_iF \bar{\nabla}^i\tau = 
- F \bar{\nabla}_i\bar{\nabla}^i\tau < 0 \ . \label{laplace} \end{equation}  

\noindent If $\xi^a$ is such that the left hand side of eq. (\ref{ojoj}) is 
positive semidefinite we conclude that 

\begin{equation} H^a\xi_a > 0  \end{equation}

\noindent at the minimum of $\tau$. Since $\xi^a$ is timelike, the mean 
curvature vector $H^a$ cannot be, and the surface cannot be trapped at the 
minimum of $\tau$.   

To return to the Friedmann model, it admits a future directed conformal 
Killing vector field for which the left hand side of eq. (\ref{ojoj}) 
is positive prior to the moment of time symmetry, and this is what we 
need in order to prove that 
%
%
%
the moment of time symmetry is indeed the trapping boundary. Through every 
point of this trapping boundary there passes a marginally 
trapped and indeed minimal closed surface. Examples 
where this is not so include open recollapsing Friedmann 
models where the trapping boundary is filled by 
minimal surfaces that are not closed. An intermediate case is 
provided by the BTZ wormhole, where minimal closed surfaces are dense in 
the trapping boundary, but do not pass through every point there \cite{brill1}. 

The minimum argument can be used to give a stronger conclusion about the 
trapping boundary. At a minimum in $\tau$ 

\begin{equation} \gamma^{ij}e^a_ie^b_j\nabla_a\xi_b = 
- F\bar{\nabla}_i\bar{\nabla}^i \tau + H^a\xi_a \ . \end{equation}

\noindent The left hand side is a projection of the second fundamental form 
of 
a hypersurface orthogonal to $\xi^a$. The Laplacian 
is positive at the minimum, and the scalar product is negative if the surface 
is trapped, so the right hand side is negative. Therefore the second fundamental 
form must have at least one negative eigenvalue if a trapped surface otherwise 
to its future touches a hypersurface in a point. In 2+1 dimensional Minkowski 
space this is easy to see: A trapped ``surface'' is 
then a spacelike curve bending downwards in inertial time, and a surface with 
positive definite second fundamental form is like a hyperboloid 
bending upwards in time. Obviously a trapped curve otherwise in its future 
cannot touch it in a point. 

\section{The self-similar Vaidya solution}
We now turn to Vaidya's spacetime, with the metric 

\begin{equation} ds^2 = - \left( 1 - \frac{2m(v)}{r}\right)dv^2 + 2dvdr 
+ r^2d\theta^2 + r^2\sin^2{\theta}d\phi^2 \ . \end{equation}

\noindent Einstein's equations read 

\begin{equation} G_{ab} = 8\pi T_{ab} = \frac{2\dot{m}}{r^2}\nabla_a v 
\nabla_bv \ . \end{equation}

\noindent The energy conditions are obeyed provided that $\dot{m} \geq 0$, where 
the dot denotes differentiation with respect to $v$. Otherwise the rate at which 
radiation comes in is at our disposal. We choose 

\begin{equation} m = \left\{\begin{array}{cll} 0 & , & v \leq 0 \\ 
\mu v & , & 0 \leq v \leq M/\mu \\ 
M & , & v \geq M/\mu \end{array} \right. \ . \label{mass} \end{equation}

\noindent  This describes a spherically symmetric shell of incoherent radiation 
entering flat spacetime from past null infinity, ending in a Schwarzschild black 
hole when the inflow stops---provided that $\mu > 1/16$, otherwise the result is a 
naked singularity \cite{HWE, Papapetrou}. Other choices of mass function have 
been studied \cite{Krishnan}. 
Our choice is special because the Vaidya region of the solution has 
a homothetic Killing vector 

\begin{equation} \eta = v\partial_v + r\partial_r \ , \label{eta} \end{equation}

\noindent whose flow lines are confined to hypersurfaces with constant $x$, where  
$x = v/r$. This extra symmetry will be quite helpful to us. We expect 
that the behaviour 
for a somewhat different mass function would be qualitatively similar but harder to 
get. The round spheres at $r = 2m$ are marginally trapped and foliate a regular 
dynamical horizon. We refer to it as the apparent horizon, because it is the apparent 
horizon in a spherically symmetric slicing of spacetime, and we do not wish to prejudge 
the issue whether there are other, non-spherically symmetric dynamical horizons present. 


In a spherically symmetric spacetime the coordinate $r$ has a meaning related to 
the area of the preferred round spheres. There is also a very special vector field 
known as the Kodama vector field $\xi$ \cite{Kodama}. In our case it is $\xi = 
\partial_v$. It is not a Killing vector 
field in general, but it does define a direction in which the area of the preferred 
round spheres is constant. It is hypersurface 
forming, and future directed and timelike outside the apparent horizon. Through 
eq. (\ref{tau}) it defines a "Kodama time" $\tau$ in this region, and this 
can be used in the minimum argument presented in section 2. A trapped surface 
sticking out of the apparent horizon at some value of $\tau$ can only reach 
higher values in its exterior. Therefore trapped surfaces at 
values of $\tau$ smaller than the smallest value $\tau_{\Sigma}$ 
assumed on the apparent horizon itself are excluded. Setting $\tau = \tau_{\Sigma}$ 
defines a spacelike hypersurface $\Sigma$, touching the apparent 
horizon just where the latter joins the event horizon. It is spherically symmetric 
and defined by a function $v = v(r)$ obeying 

\begin{equation} \frac{dv}{dr} = \frac{1}{1 - \frac{2m}{r}} \ . \end{equation}

\noindent This is easily solved for in terms of elementary functions in the 
self-similar case \cite{Papapetrou}. In the flat region constant Kodama time 
means constant intertial time, and the hypersurface $\Sigma$ will enter the 
flat region if and only if $\mu > 1/8$. 

Is it possible that $\Sigma$ is the trapping boundary we look for? In fact 
it cannot be, since the eigenvalues of its second fundamental form are 
$(k_1, k_2,k_2) = (k_1, 0, 0)$ where $k_1 > 0$. But we have already argued 
that such a hypersurface cannot be touched by marginally trapped surfaces. 
There must be a region to the future of $\Sigma$ which is free of trapped 
surfaces, but at least we have constrained the trapping boundary from below. 
We have also proved that the latter must be spacelike close to the event horizon, 
since it is squeezed from below by the spacelike $\Sigma$.    

\section{Tongues sticking out of the apparent horizon}
Is it possible that the trapping boundary coincides with the apparent 
horizon? In fact no \cite{IBJS}. 
An argument due to Galloway and Wald shows that marginally 
trapped round spheres in a spherically symmetric dynamical horizon can 
always be perturbed in such a way that they extend partly outside it, and 
such that the perturbation causes them to become trapped \cite{Wald}. But 
their argument does not show how far below the apparent horizon these trapped 
surfaces extend. Here we will present a---non-optimal---construction which 
allows us to find closed trapped 
surfaces extending a finite distance away from the apparent horizon. 

Our surfaces are defined as cross sections of a cone, by means of the two 
equations  

\begin{equation} v = \frac{k}{2\mu}r - v_0 \ , \hspace{8mm} v = \frac{1}{2\mu}r 
+ a(\theta ) \ . \label{surface} \end{equation}

\noindent When $a = 0$ this surface sits on the apparent horizon. When $|a|$ is 
large the surface may extend into the Schwarzschild or Minkowski regions of the 
solution, but we stick to the above definition---it defines a smoothly 
embedded surface in ${\bf R}^4$, although its first fundamental form will not be 
smooth and its null expansions will jump at the boundaries between different 
regions. 

The condition for outer trapping is the somewhat lengthy inequality   

\begin{eqnarray} N H(k_+) = - 2\alpha r^3 +2r^3
\left(1-\frac{2m}{r}\right) 
+ \frac{\alpha k - 2\mu}{k-1}r^2(a^{\prime \prime} + a^\prime \cot{\theta}) + 
\nonumber \\ 
\ \nonumber \\
+ \frac{a^{\prime 2}}{(k-1)^2} \left( 4\mu km + 8\mu^2r + \alpha k^2m - 8\mu \alpha kr 
- k^2\dot{m}r \right) + \hspace{10mm} \nonumber \\
\ \label{Trapm} \\ 
+ \frac{a^{\prime 2}}{(k-1)^2} \left(1-\frac{2m}{r}\right) \left( 4\mu kr + \alpha k^2r - k^2m 
- r\left(1-\frac{2m}{r}\right)k^2\right) + \hspace{10mm} \nonumber \\ 
\nonumber \\
+ \frac{ka^{\prime 2}}{(k-1)^2}\left( 4\mu - k\left( 1-\frac{2m}{r}\right) \right) 
\frac{\alpha k - 
2\mu}{k-1}a^\prime \cot{\theta} < 0 \ , \nonumber \end{eqnarray} 

\noindent where $H(k_+)$ is the outer null expansion, $N$ is a positive normalisation 
factor, and 

\begin{equation} \alpha = \frac{2\mu}{k} + \frac{(k-1)^2r^2}{k^2a^{\prime 2}}\left( 
1 - \sqrt{1 + \frac{4\mu ka^{\prime 2}}{(k-1)^2r^2} - \left( 1 - \frac{2m}{r}\right) 
\frac{k^2a^{\prime 2}}
{(k-1)^2r^2}}\right) \ . \end{equation}   

\noindent To simplify the calculation we set 

\begin{equation} k = 1 + l(l+1) \ , \hspace{8mm} 
a(\theta ) = a_0 + a_lP_l(\cos{\theta}) \ , \end{equation}   

\noindent where $P_l$ is a Legendre polynomial. Together with eqs. (\ref{surface}) this 
defines a set of ``tongues'' that---as we will see---stick partly out of the apparent 
horizon. To first order in $a$ the trapping condition becomes simply $a_0 > 0$, and the 
tongue extends partly outside the apparent horizon if there is a $\theta$ such that 
$a(\theta ) < 0$. This is the perturbation considered by Galloway and Wald \cite{Wald}. 
 
Our aim is now to adjust the parameters $a_0$ and $a_l$ for every given $v_0$ in 
such a way that the extent to which the tongue sticks out of the horizon is 
maximised. To deal with the inequality (\ref{Trapm}) we 
had to fall back on Mathematica. General considerations about when 
the cone becomes tangential to hypersurfaces of constant Kodama time show that 
the possible extent of the tongue is larger the smaller $l$ is, so we confined 
ourselves to $l = 1$ and $l = 2$. Then we investigated the two cases 
$\mu = 1/2$ and $\mu = 2$ in detail and found---for the case when the tongues 
remain in the Vaidya region of spacetime---that the most stringent condition 
from the inequality comes at the tips of the tongue, where $a^\prime = 0$. 
But once this is so the inequality (\ref{Trapm}) is easily dealt with 
analytically. For $m = \mu v$, $l = 1$, the 
trapping conditions at the tips hold as long as 

\begin{equation} \mu = \frac{1}{2} \ : \hspace{8mm} \ 0.856 < x = \frac{v}{r} 
< 1.309 \ \ \end{equation}

\begin{equation} \mu = 2 \ : \hspace{8mm} - 0.417 < x = \frac{v}{r} < 0.75  
\ . \end{equation}

\noindent This is then how far a maximally extended closed trapped tongue extends. 
The inner trapping condition holds throughout the allowed region. Thus the region 
occupied by trapped tongues is bounded by 
hypersurfaces of constant $x$. Given that the trapping condition becomes 
critical at the tips of the tongues this could have been predicted, because 
it is known that if the outer null expansion vanishes at a point it remains 
zero if the surface is moved by a homothety \cite{Carrasco}. In the 
self-similar Vaidya solution this is generated by the vector field $\eta$, 
see eq. (\ref{eta}).   

\begin{figure}
\begin{center}
\includegraphics[width=3.1in]{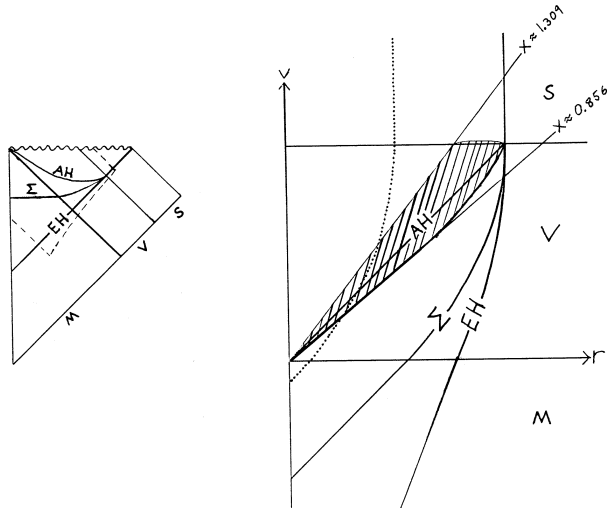}
\end{center}
\caption{\label{label}To the left is a Penrose diagram of a Vaidya spacetime. The region inside the dashed curve is shown as a $v-r$-diagram to the right, and includes a part of the event horizon (EH), the spacelike part of the apparent horizon (AH), and the hypersurface $\Sigma$ below which no trapped surfaces can extend. The tongues considered in the text are confined between two lines of 
constant $x$, and can extend only a little bit into the Schwarzschild region 
S. The trapped surfaces discussed 
earlier \cite{IBJS} are to the left of the dotted curve. The figure is for $\mu = 1/2$ only.} 
\end{figure}

For the two values of $\mu$ that we investigated in detail we also 
found that the intrinsic curvature $R$ of the tongue obeys $R > 0$ throughout 
the region where it is trapped, thus disproving a conjecture by Hayward 
\cite{Sean}.   

So far we have avoided the complication that the tongues may extend into 
the Minkowski or Schwarzschild regions, where the mass function is different. 
This must now be dealt with. It is no longer true that the trapping condition 
is always at its most stringent at the tips of the tongue. After a lengthy 
Mathematica calculation we found, however, that the 
upper Schwarzschild tip is usually the place where trapping fails first. 
The result is that the region into which the tongues can extend shrinks---as 
indeed it must, since its boundary must reach the point where the 
dynamical horizon meets the event horizon. See Fig. 1 for a summary of the 
calculation in the $\mu = 1/2$ case. For $\mu = 2$ the tongues can extend 
into the flat region.

It is not the case though that the Schwarzschild region makes life more 
difficult for all kinds of closed trapped surfaces. Indeed the flat and 
Vaidya region has locally trapped surfaces in the equatorial plane 
$\theta = \pi/2$. Topologically they are open disks meeting the Schwarzschild 
boundary in a circle. If they are 
carefully adjusted, and do not extend to large values of $r$, it is possible 
to close them in the Schwarzschild region while keeping them trapped, so 
the Schwarzschild region in a sense creates closed trapped surfaces 
that could not exist without it \cite{IBJS}.    

\section{So where is the trapping boundary?}
A large part of the no-man's land between the 
dynamical horizon and the event horizon has now been occupied by the trapped 
region, but we have not been able to pin down its boundary precisely. 
Some things are clear: in a spherically symmetric spacetime the trapping 
boundary must itself be spherically symmetric, it cannot contain any 
marginally trapped surfaces, it cannot have a positive definite 
second fundamental form, and it must be spacelike close to the event 
horizon because it is squeezed from below by the spacelike hypersurface 
$\Sigma$. From our picture for $\mu = 1/2$ it is tempting to conjecture 
that a part of the boundary sits at $x = $ constant. This part 
would be determined entirely by the behaviour of trapped surfaces inside 
the self-similar Vaidya region. 
But for $\mu = 2$ the trapping boundary 
is determined largely by what happens inside the Schwarzschild region---not 
by the local physics in the Vaidya region.

Our calculations have not addressed the question to what extent the dynamical 
horizon is unique. This also remains as an interesting question.
  
\section*{Acknowledgements}
IB thanks Greg Galloway and Catherine Williams for discussions, and 
the Swedish Research Council VR for support. JMMS is supported by grants 
FIS2004-01626 (MICINN) and GIU06/37 (UPV/EHU). J\AA \ and IB found 
the Spanish Relativity Meeting as stimulating as ever.

\end{document}